\documentclass[9pt,conference]{IEEEtran}

\usepackage[preprint]{waspaa25}
\usepackage{bm} 
\usepackage{multirow, booktabs, bbding, enumerate, enumitem, subcaption, xcolor}%

\title{SpecMaskFoley: Steering Pretrained Spectral Masked Generative Transformer Toward Synchronized Video-to-audio Synthesis via ControlNet}

\name{
      Zhi Zhong$^{1*}$,
      Akira Takahashi$^{1*}$,
      Shuyang Cui$^{1}$, 
      Keisuke Toyama$^{1}$,
      Shusuke Takahashi$^{1}$,
      Yuki Mitsufuji$^{1, 2}$
      \thanks{$^*$Equal contribution.}}
\address{$^{1}$Sony Group Corporation, Japan \;
$^{2}$Sony AI, USA\
}

\begin{document}

\maketitle
\begin{abstract}
Foley synthesis aims to synthesize high-quality audio that is both semantically and temporally aligned with video frames. Given its broad application in creative industries, the task has gained increasing attention in the research community.
To avoid the non-trivial task of training audio generative models from scratch, adapting pretrained audio generative models for video-synchronized foley synthesis presents an attractive direction. 
ControlNet, a method for adding fine-grained controls to pretrained generative models, has been applied to foley synthesis, but its use has been limited to handcrafted human-readable temporal conditions.
In contrast, from-scratch models achieved success by leveraging high-dimensional deep features extracted using pretrained video encoders.
We have observed a performance gap between ControlNet-based and from-scratch foley models.
To narrow this gap, we propose SpecMaskFoley, a method that steers the pretrained SpecMaskGIT model toward video-synchronized foley synthesis via ControlNet.
To unlock the potential of a single ControlNet branch, we resolve the discrepancy between the temporal video features and the time-frequency nature of the pretrained SpecMaskGIT via a frequency-aware temporal feature aligner,
eliminating the need for complicated conditioning mechanisms widely used in prior arts. 
Evaluations on a common foley synthesis benchmark demonstrate that SpecMaskFoley could even outperform strong from-scratch baselines, substantially advancing the development of ControlNet-based foley synthesis models.
Demo page: \url{https://zzaudio.github.io/SpecMaskFoley_Demo/}.
\end{abstract}

\vspace{-2.5mm}
\section{Introduction}
\label{sec:intro}
\vspace{-0.5mm}
Text-to-audio (TTA) synthesis targets at synthesizing realistic sound events by natural language prompts \cite{kreuk2022audiogen, huang2023makeanaudio2, comunita2024specmaskgit, saito2024soundctm, evans2025sa-open}. 
In spite of the success of TTA systems in terms of sound quality as well as the semantic alignment between audio and text, the use of such systems in \textit{foley} synthesis is very limited. 
Foley synthesis aims to synthesize audio that is not only semantically but also \textit{temporally} aligned with video frames \cite{cheng2024mmaudio}. Automated foley synthesis has broad impact to creative industries, thus has gained arising attention in the research community.

Temporal alignment for foley synthesis can be learned via the joint generative modeling of audio-visual pairs (\cite{ruan2023mmdiffusion, yang2024visualechoes}), though the training can be expensive. 
Although there have been attempts to achieve temporal control of audio objects using text prompts \cite{huang2023makeanaudio2}, 
the mainstream is to explicitly condition audio generative models with temporal features that are synchronized with the video.
To avoid the non-trivial task of training an audio generative model from scratch, steering \textit{pretrained} audio generative models toward foley synthesis presents an attractive direction. 
Adapting or aligning pretrained audio and video latent spaces has been investigated (\cite{wang2024v2amapper, xing2024seeingandhearing}), but the temporal alignment between video and audio is poor \cite{cheng2024mmaudio}.

\begin{figure}[tbp]
  \centering
  \includegraphics[width = 0.48\textwidth]{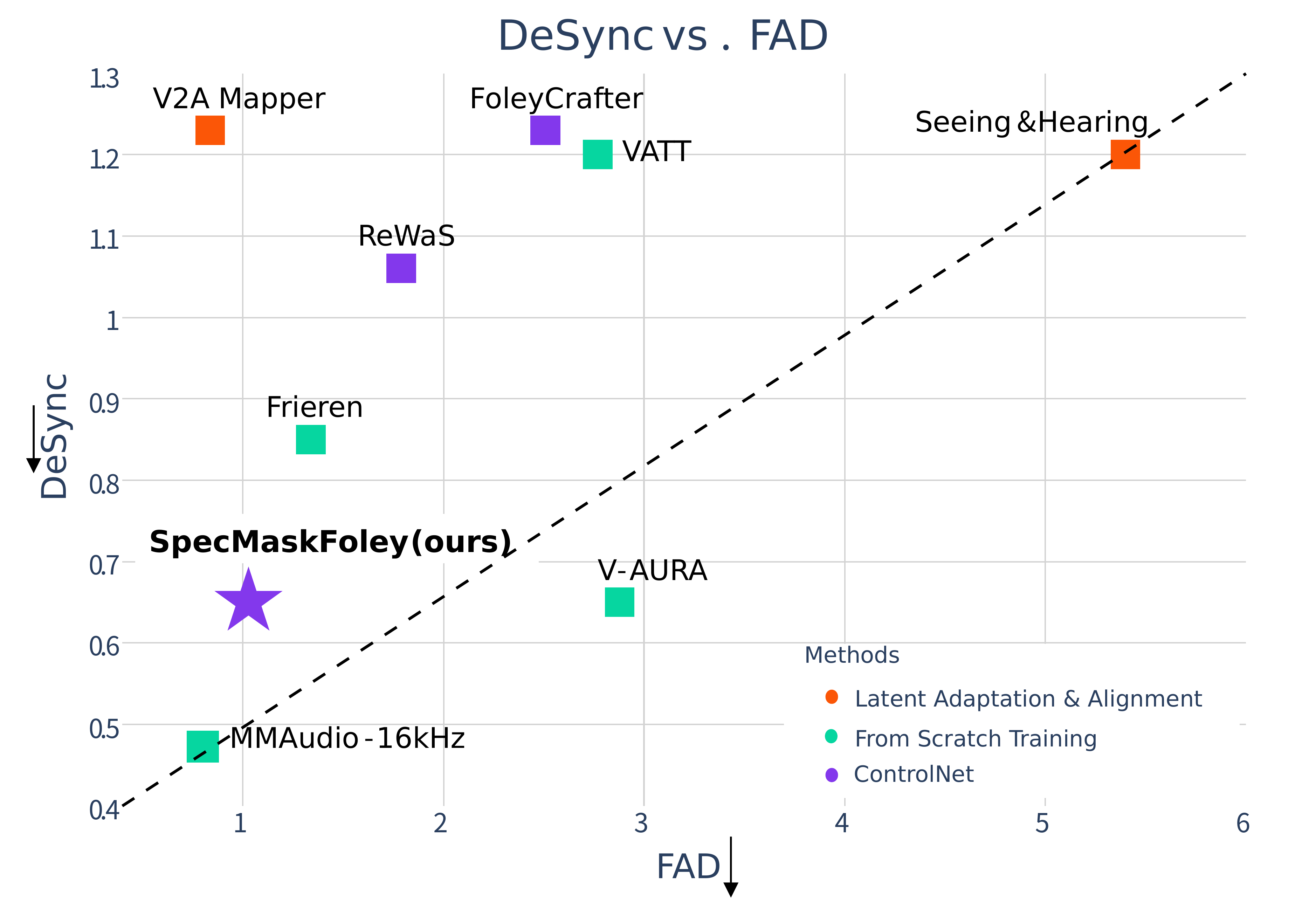}
  \vspace{-0.5mm}
  \caption{Audio synthesis quality (FAD \cite{kilgour2018fad}) and audio-video temporal alignment \cite{iashin2024synchformer} of different methods. The proposed SpecMaskFoley achieves competitive scores in both axes without from-scratch training.}
  \label{fig:fad_vs_desync}
  \vspace{-2mm}
\end{figure}
%

ControlNet (\cite{zhang2023controlnet, zhao2023unicontrolnet}), originally designed to add \textit{spatial} controls to pretrained image generative models, has been introduced to foley synthesis to explicitly inject handcrafted human-readable \textit{temporal} conditions into pretrained audio generative models (\cite{jeong2024rewas, zhang2024foleycrafter}). 
Attempts also have been made to train foley synthesis models from-scratch with high-dimensional \textit{deep features} extracted using pretrained video encoders \cite{liu2024vatt, viertola2025vaura, wang2024frieren, cheng2024mmaudio}. With advanced video encoders such as masked Transformers \cite{huang2023mavil} and contrastive learning models \cite{luo2023difffoley, iashin2024synchformer}, from-scratch models like Frieren \cite{wang2024frieren} and MMAudio \cite{cheng2024mmaudio} have brought foley synthesis to a new level.

As shown in Fig. \ref{fig:fad_vs_desync}, a notable performance gap can be observed between from-scratch models and ControlNet models, indicating that the effective use of pretrained audio generative models in the foley synthesis task is yet to be explored.
To narrow this gap, we propose \textit{SpecMaskFoley}, a method that steers the pretrained SpecMaskGIT model \cite{comunita2024specmaskgit} toward video-synchronized foley synthesis via ControlNet. 
Our contributions lie in the following aspects.
\textbf{(1) Competitive foley synthesis performance as a ControlNet-based method.} 
SpecMaskFoley outperforms not only prior ControlNet-based foley models \cite{jeong2024rewas, zhang2024foleycrafter}, but also strong from-scratch baselines, including Auto-Regressive model (AR) \cite{viertola2025vaura}, MaskGIT model \cite{liu2024vatt}, and flow-matching model \cite{wang2024frieren}, in a widely used foley synthesis benchmark.  
\textbf{(2) Simplicity in neural network architecture.} Prior foley synthesis methods tend to combine multiple conditioning mechanisms for better video-audio alignment, \textit{e.g.}, the combination of cross-attentions adaptors and ControlNet \cite{zhang2024foleycrafter}. Nevertheless, SpecMaskFoley utilizes a single ControlNet branch with a Frequency-aware Temporal feature Aligner (FT-Aligner) to effectively adapt the temporal video features to our time-frequency ControlNet branch. The success of our simple design underscores the potential of ControlNet to process complex, high-dimensional features, paving the way to future extending Controlnet to challenging audio generation tasks. 

%
\begin{figure*}[t]
  \centering
  \includegraphics[width = 0.7\textwidth]{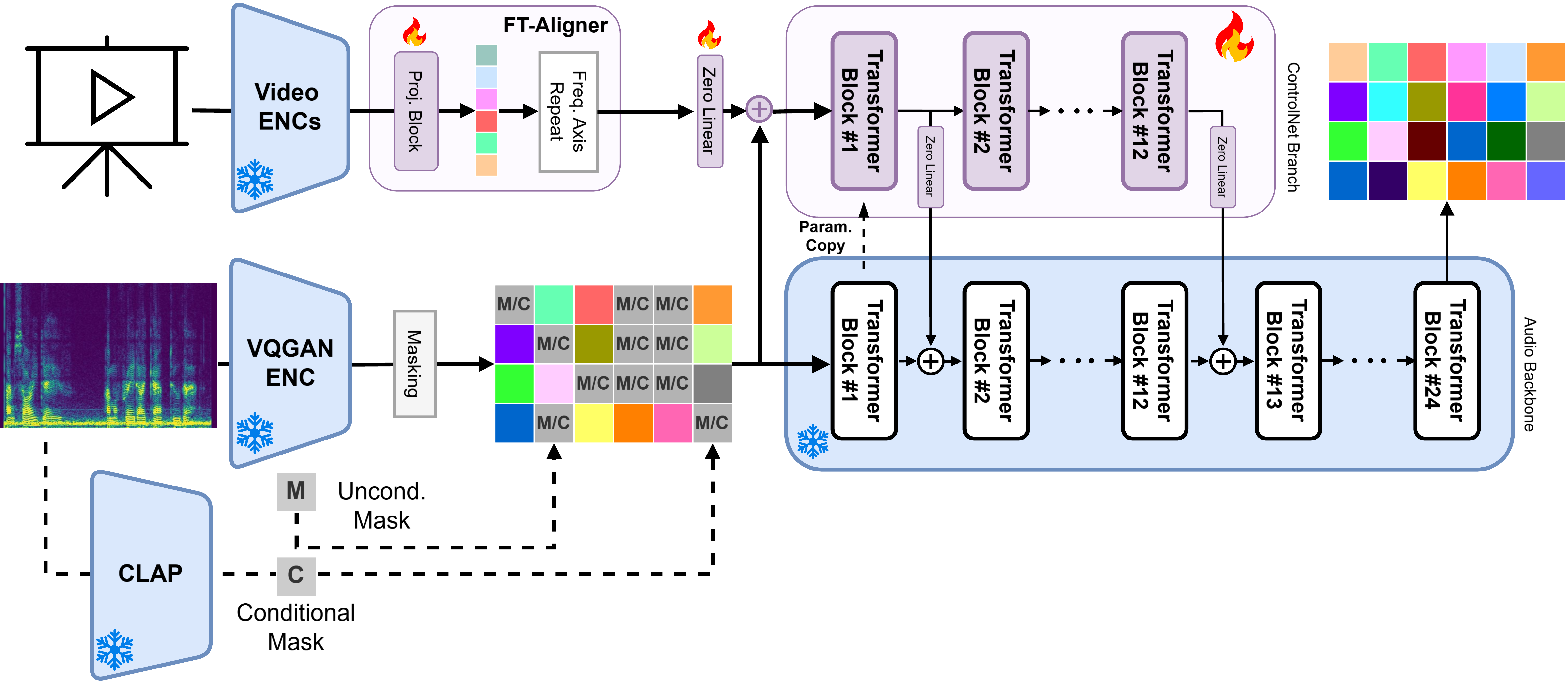}
  \vspace{-1.5mm}
  \caption{Overview of SpecMaskFoley. Ice icons: frozen modules. Fire icons: trainable modules. A CLAP embedding is treated as a conditional mask $C$ following \cite{comunita2024specmaskgit} to condition the audio backbone with audio prompts during training and text prompts during inference.}
  \vspace{-1.5mm}
  \label{fig:SpecMaskFoley}
\end{figure*}
%
\vspace{-1.5mm}
\section{Related Works}
\label{sec:related_works}
\vspace{0.5mm}
While significant progress has been made in TTA synthesis \cite{kreuk2022audiogen, huang2023makeanaudio2, comunita2024specmaskgit, saito2024soundctm, evans2025sa-open}, it is still difficult to achieve fine-grained temporal control using natural language prompts \cite{huang2023makeanaudio2}. 
The joint generative modeling of synchronized visual and audio data leads to an aligned latent space, such as in MMDiffusion \cite{ruan2023mmdiffusion} and VisualEchoes \cite{yang2024visualechoes}. However, jointly modeling multiple modalities from scratch is non-trivial. 
The adaptation or alignment of pretrained audio and video latent spaces has been explored to mitigate the burden of from-scratch training (\cite{wang2024v2amapper, xing2024seeingandhearing}), but resulted in poor temporal alignment between video and audio \cite{cheng2024mmaudio}.

ControlNet (\cite{zhang2023controlnet, zhao2023unicontrolnet}), originally designed to add spatial controls to pretrained image generative models, has been introduced to foley synthesis to explicitly inject handcrafted human-readable temporal conditions into pretrained audio generative models. 
Typical handcrafted features include the energy curves (\cite{gramaccioni2024stablev2a, jeong2024rewas}), onset timestamps \cite{zhang2024foleycrafter}, and binarized CLIP stamps \cite{zhang2025smoothfoley}. 
FoleyCrafter \cite{zhang2024foleycrafter} further combines ControlNet with parallel cross-attention adaptors to make use of CLIP \cite{radford2021clip} visual features.

Parallel to the investigation of ControlNet-based models, attempts have been made to train foley synthesis models from scratch. 
From-scratch models often take the advantage of various high-dimensional deep features extracted using pretrained visual encoders.
For example, VATT \cite{liu2024vatt} trains a MaskGIT (\cite{chang2022maskgit, comunita2024specmaskgit}) model controlled by eva-CLIP \cite{sun2023evaclip} visual features. 
V-AURA trains an AR Transformer (\cite{kreuk2022audiogen, liao2024sonido}) with the deep features extracted by Synchformer \cite{iashin2024synchformer}.
Both VATT and V-AURA are trained upon a discrete latent space created by 1-D VQ-GANs (\cite{defossez2022encodec, kumar2023dac}).
Frieren \cite{wang2024frieren} trains a flow-matching model upon the latent space created by a 1D-VAE-GAN (\cite{huang2023makeanaudio2}) conditioned by deep features from both contrastive learning (\cite{luo2023difffoley}) and masked Transformer models (\cite{huang2023mavil}). 
MMAudio \cite{cheng2024mmaudio} brought the foley synthesis task to a new level. Similar to Frieren, MMAudio is a flow-matching model built upon a 1D-VAE-GAN, while taking the advantage of the Synchformer features (\cite{viertola2025vaura, iashin2024synchformer}) and CLIP visual features \cite{radford2021clip}.

A method that more effectively leverages pretrained audio generative models is likely to narrow the notable performance gap observed between from-scratch and ControlNet-based models.

\vspace{1mm}
\section{SpecMaskFoley}
\label{sec:proposed_method}
\vspace{0.25mm}
As shown in Fig.\ref{fig:SpecMaskFoley}, SpecMaskFoley consists of a pretrained SpecMaskGIT \cite{comunita2024specmaskgit} TTA model as the audio backbone, ControlNet branch, and an FT-Aligner to adapt temporal deep video features to our time-frequency ControlNet implementation.
\subsection{SpecMaskGIT}
\label{ssec:specmaskgit}
SpecMaskGIT features a distillation-free approach to efficient high-quality TTA synthesis \cite{comunita2024specmaskgit}. 
The efficiency, effectiveness and flexibility of this method are based on: (1) A highly compressed latent space created by a 2-D SpecVQGAN \cite{iashin2021SpecVQGAN}; (2) the light-weight ViT backbone trained with the Masked Language Modeling task (\cite{he2022mae, zhong2023ExtendedAudioMAE, yang2024visualechoes}) upon the latent space of SpecVQGAN; (3) the use of parallel iterative synthesis, \textit{i.e.}, MaskGIT  sampler, for fast inference (\cite{chang2022maskgit, li2023mage}).

We chose to use SpecMaskGIT as the backbone of SpecMaskFoley for the following reasons. 
First, SpecMaskGIT has been reported \cite{comunita2024specmaskgit} to be competitive in a TTA benchmark -- on par with AudioLDM \cite{liu2023audioldm}, a TTA model widely used in foley synthesis research (\cite{wang2024v2amapper, xing2024seeingandhearing, jeong2024rewas, zhang2024foleycrafter}) -- while being more efficient in inference than prior arts. 
Second, the lightweight and efficient nature of SpecMaskGIT enables fast concept verification and iterative try-and-error exploration with academic level of resources. 
Finally, MaskGIT methods showed promising results in foley synthesis (\cite{yang2024visualechoes, liu2024vatt}).
\subsection{Deep Video Features}
\label{ssec:visual_feature}
Instead of using handcrafted features as with prior ControlNet methods (\cite{jeong2024rewas, zhang2024foleycrafter}), we adopt the same deep video features as in \cite{cheng2024mmaudio, viertola2025vaura} to our ControlNet branch:
A 25 Hz high-frame-rate video feature extracted using Synchformer \cite{iashin2024synchformer} and an 8 Hz semantic feature extracted using CLIP \cite{radford2021clip}. 
Both features are 1-D sequences of high dimensionality. 
Although the 8 Hz CLIP features have been used in MMAudio \cite{cheng2024mmaudio},
we hypothesize that CLIP features present more semantic rather than temporal synchronization information, thus can be averaged across all frames to form a global semantic condition. 
This global condition is directly added to Synchformer features, resulting in a 1-D temporal features of shape [$t$, $d$], where $t$ is the number of temporal frames of the deep feature, and $d$ is the dimension.
\subsection{ControlNet and FT-Aligner}
\label{ssec:controlnet}
As shown in Fig.\ref{fig:SpecMaskFoley}, we implement a Transformer version of ControlNet similar to \cite{chen2024pixart, ciranni2024stableaudiocontrol, hou2025stableediting}, which initializes the ControlNet branch with the pretrained model weights and then connects the ControlNet branch to the backbone via zero-initialized linear layers. Although this design was presented for continuous Diffusion Transformers \cite{chen2024pixart, ciranni2024stableaudiocontrol, hou2025stableediting}, we found it works well in our discrete model. Note that the only training loss in SpecMaskFoley is cross entropy loss, the standard training target for discrete models (\cite{chang2022maskgit,li2023mage,comunita2024specmaskgit,yang2024visualechoes}).

A major challenge to control the pretrained SpecMaskGIT model with \textit{1-D} temporal features is the \textit{2-D} time-frequency nature of SpecMaskGIT. 
SpecMaskGIT works in a 2-D latent space, where each token represents a 16-by-16 time-frequency patch in the original Mel-spectrogram.
To achieve effective temporal alignment, it is essential to inject the \textit{identical} temporal feature to all tokens belonging to the same time frame.
%
To this end, we propose a Frequency-aware Temporal feature Aligner (FT-Aligner) to address this challenge. 
Assuming the audio token sequence sent into SpecMaskGIT has a shape of [$F$, $T$, $D$], denoting the number of tokens along the frequency axis, number of tokens along the temporal axis, and dimensionality for each token. 
For 1-D temporal video features of shape [$t$, $d$] described in Sec.\ref{ssec:visual_feature}, we use a Projection Block containing a 1-D conv layer followed by an adaptive average pooling to downsample the 1-D features to [1, $T$, $D$], 
then \textit{repeat} the sequence along the frequency axis, resulting 2-D feature embeddings with the shape of [$F$, $T$, $D$] that preserves original 1-D temporal information.
We empirically found this FT-Aligner facilitates the convergence in training. Without this careful feature alignment, we would not be able to train SpecMaskFoley.

\subsection{Multi Classifier-Free Guidance}
\label{ssec:multicfg}
Classifier-Free Guidance (CFG \cite{ho2022classifier}) has been used to improve TTA synthesis quality by balancing between diversity and audio-text alignment (\cite{comunita2024specmaskgit, chang2022maskgit, chang2023muse}). 
Inspired by StemGen \cite{parker2024stemgen}, we extend the original single-condition CFG formula to two conditions, effectively using both video and text conditional features when available:
\begin{equation}
\label{eq:cfg}
    \ell_\mathrm{foley} = \ell_\mathrm{uncond} + t[(\ell_\mathrm{text\&video} - \ell_\mathrm{uncond}) + (\ell_\mathrm{video} - \ell_\mathrm{uncond})],
\end{equation} 
where $\ell_\mathrm{uncond}$ denotes the logits gained from the audio backbone without the CLAP conditioning, $\ell_\mathrm{video}$ denotes logits obtained from conditioning SpecMaskFoley with deep video features via ControlNet but without CLAP feautures in the backbone, $\ell_\mathrm{text\&video}$ denotes logits earned by conditioning SpecMaskFoley with both CLAP text features and deep video features simultaneously, and $t$ denotes the CFG scale.
We do not use logits gained by conditioning the audio backbone with CLAP text features, and we also randomly replace the CLAP conditioning with an unconditonal mask (shown in Fig.\ref{fig:SpecMaskFoley}) for 90\% of the training steps, as we found text prompts alone are less important in Sec.\ref{ssec:ablation_discussion}.

%
\begin{table*}[ht]
\centering
\caption{Benchmarking on VGGSound test set. AR.: Auto-regressive. Mask.: MaskGIT. Diff: Diffusion and flow matching. Bold: best score. Underline: the second and third best scores. Inference time is computed on a H100 GPU with batch size 1 for one 10-second clip.}
\label{tab:res_vgg_all}
\resizebox{\linewidth}{!}{
\begin{tabular}{ccccc|ccc|c|c|c}
\toprule
    \multirow{2}*{Method} & \multirow{2}*{Params$\downarrow$} & \multirow{2}*{Type} & \multirow{2}*{Pretrained Backbone} & \multirow{2}*{Video features} & \multicolumn{3}{c|}{Audio Synthesis} & AV Semantic & AV Sync. & \multirow{2}*{Infer. Time (s)$\downarrow$}\\
     &  &  &  &  & FD$\downarrow$ & FAD$\downarrow$ & KL$\downarrow$ & IB Similarity$\uparrow$ & DeSync (s)$\downarrow$ & \\
\midrule
    \textit{Latent Adaptation \& Alignment} &&&&&&&&&& \\
    Seeing \& Hearing \cite{xing2024seeingandhearing} & 415M & Diff. & AudioLDM \cite{liu2023audioldm} & ImageBind \cite{girdhar2023imagebind} & 219 & 5.40 & 2.3 & \textbf{34.0} & 1.20 & 14.6 \\
    V2A Mapper \cite{wang2024v2amapper} & \underline{230M} & Diff. & AudioLDM \cite{liu2023audioldm} & CLIP \cite{radford2021clip} \& CLAP \cite{laion2023clap} & \underline{84.6} & \underline{0.84} & 2.56 & 22.6 & 1.23 & - \\
\midrule
    \textit{ControlNet + Handcrafted Features} &&&&&&&&&& \\
    ReWaS \cite{jeong2024rewas} & 620M & Diff. & AudioLDM \cite{liu2023audioldm} & Energy Curve & 141 & 1.79 & 2.82 & 14.8 & 1.06 & 16.0 \\
    FoleyCrafter \cite{zhang2024foleycrafter} & 1.22B & Diff. & AudioLDM \cite{liu2023audioldm} & Onset timestamps & 140 & 2.51 & 2.23 & 25.7 & 1.23 & \underline{1.7} \\
\midrule
    \textit{From Scratch Training} &&&&&&&&&& \\
    VATT \cite{liu2024vatt} & - & Mask & - & eva-CLIP \cite{sun2023evaclip} & 132 & 2.77 & \textbf{1.41} & 25.0 & 1.20 & - \\
    V-AURA \cite{viertola2025vaura} & 695M & AR & - & Synchformer \cite{iashin2024synchformer} & 218 & 2.88 & 2.07 & \underline{27.6} & \underline{0.65} & 16.6 \\
    Frieren \cite{wang2024frieren} & \textbf{160M} & Diff. & - & CAVP \cite{luo2023difffoley} \& MAViL \cite{huang2023mavil} & \underline{106} & 1.34 & 2.86 & 22.8 & 0.85 & - \\
    MMAudio-16kHz \cite{cheng2024mmaudio} & \textbf{160M} & Diff. & - & Synchformer \cite{iashin2024synchformer} \& CLIP \cite{radford2021clip} & \textbf{70.2} & \textbf{0.79} & \underline{1.59} & \underline{29.1} & \textbf{0.48} & \underline{1.23} \\
\hline
    \textit{ControlNet + Deep Features} \\
    SpecMaskFoley (ours) & 300M & Mask & SpecMaskGIT \cite{comunita2024specmaskgit} & Synchformer \cite{iashin2024synchformer} \& CLIP \cite{radford2021clip} & 109 & \underline{1.03} & \underline{1.76} & 26.4 & \underline{0.65} & \textbf{0.47} \\
\bottomrule
    \end{tabular}}
\end{table*}
\vspace{1mm}
\section{Experiments}
\label{sec:exp}
\vspace{-1mm}
\subsection{Datasets}
\label{ssec:dataset_metrics}
The pretraining of the TTA backbone in SpecMaskFoley was conducted on the sum of the unbalanced and balanced subset of AudioSet \cite{gemmeke2017audioset}, a dataset that has been widely used in general audio representation learning (\cite{koutini2021patchout, zhong2023ExtendedAudioMAE}) and generative modeling (\cite{huang2023makeanaudio2, liu2023audioldm})  due to its massive amount of audio clips as well as its diversity in sound sources and recording environments.

The ControlNet branch of SpecMaskFoley is trained on VGGSound \cite{chen2020vggsound}, the only audio-visual dataset used in this study, which contains around 500 hours sounding video clips. 
On top of its synchronized audio-video pairs, video clips in VGGSound come with tags from a 310-class taxonomy.
We follow the data split and preprocessing pipeline in MMAudio \cite{cheng2024mmaudio}, in which the train set contains around 180K 10-second video clip. However, we do not truncate the videos to 8s, as our audio backbone was pretrained with 10-second audio clips.

Following common practice in ReWaS\cite{jeong2024rewas}, VATT\cite{liu2024vatt}, FoleyCrafter\cite{zhang2024foleycrafter}, and MMAudio\cite{cheng2024mmaudio}, we concatenate tags of the test set as the text input to SpecMaskFoley during evaluation.

\subsection{Implementation Details}
\label{ssec:implementation_details}
For the TTA backbone, we use the official checkpoint of SpecMaskGIT \cite{comunita2024specmaskgit}, which was pretrained on the AudioSet. 
More details of pretraining can be found in \cite{comunita2024specmaskgit}.
The standard Mel-spectrogram transform from vocoders \cite{kong2020hifigan} is used, which transforms 10-second audio clip at the sampling rate 22.05kHz into 848 frames with 80 Mel bins. 
The Mel-spectrogram is further tokenized using SpecVQGAN with a 16-by-16 downsampling factor, resulting in a 2-D token map of [$F = 5, T = 53$], while each token in the map is represented by a 256-dimension embedding from a 10bit codebook.
Each 10-second video clip is processed by Synchformer \cite{iashin2024synchformer} into 240 feature frames, and by CLIP \cite{radford2021clip} into 80 feature frames respectively. As mentioned in Sec.\ref{ssec:visual_feature}, these CLIP feature frames are then globally averaged.

The audio backbone in SpecMaskFoley employs 24 Transformer blocks, in which the attention dimension is $D = 768$ with 8 heads and the feedforward dimension is 3072, resulting in around 170M parameters. 
We copy the first 12 Transformer blocks, \textit{i.e.}, half of the audio backbone, to initialize the ControlNet branch in SpecMaskFoley. The total number of \textit{trainable} parameters is around 126M. 
To align deep video features with the shape of the 2-D audio token map, the proposed FT-Aligner first downsamples the aforementioned deep features to [$F = 1, T = 53$, $D = 768$] then repeats the features 5 times along the frequency axis. 

 We trained the ControlNet branch for 140K steps on a \textit{single} A6000 GPU. Following common practice (\cite{koutini2021patchout, zhong2023ExtendedAudioMAE}), we employ a linear warmup for the first 2.8 K steps then a cosine annealing of the learning rate (LR) for the remaining training. The batch size is set to 64, the base LR is set to 1e-3. The LR equates to the base LR times the batch size divided by 256 (\cite{li2023mage, comunita2024specmaskgit}).

Unless denoted, we use the multi-CFG described in Sec.\ref{ssec:multicfg} with the CFG scale linearly increasing from 0 to 3 across the 12 inference steps with the Gumbel temperature (\cite{comunita2024specmaskgit}) set to 9.0 during evaluation.

\subsection{Metrics}
\label{ssec:implementation_details}
We use the av-benchmark \footnote{\url{https://github.com/hkchengrex/av-benchmark}} to evaluate the quality of foley synthesis from the following aspects:
\textbf{Audio synthesis quality}. 
Following common practice (\cite{kreuk2022audiogen, huang2023makeanaudio2, comunita2024specmaskgit, saito2024soundctm, cheng2024mmaudio}), we compute the Frechet Distance (FD) and Kullback–Leibler (KL) distance on the basis of PaSST \cite{koutini2021patchout}, a 32 kHz audio classifier.
The 16 kHz VGGish classifier \cite{hershey2017vgg} is also used for FD (denoted as ``FAD"). 
Note that we exclude PANN \cite{kong2020panns} from the metric computation, as it has been reported as not being robust in some scenarios \footnote{\url{https://github.com/haoheliu/audioldm_eval}}.
\textbf{Audio-video semantic matching.} The semantic similarity between the input video and the generated foley audio is evaluated by the cosine similarity between ImageBind \cite{girdhar2023imagebind} video and audio features, denoted as “IB similarity”.
\textbf{Audio-video synchronization}. The synchronization between input video and generated audio is measured by DeSync (\cite{cheng2024mmaudio, iashin2024synchformer}) in seconds.
\section{Results}
\label{sec:res}
\subsection{VGGSound Benchmarking}
\label{ssec:vggsound_res}
Benchmarking results on VGGSound are presented in Tab. \ref{tab:res_vgg_all}, in which the top-3 scores are highlighted. 

Several observations can be made. First, MMAudio \cite{cheng2024mmaudio} ranks in the top-3 among all metrics, which sets a high standard for foley syntheis and indicates room of improvement for SpecMaskFoley.
Second, the proposed SpecMaskFoley outperforms those latent adaptation methods in DeSync while maintaining competitive audio synthesis quality, indicating the effectiveness of using ControlNet to enhance the video-audio synchronization.
Next, SpecMaskFoley outperforms previous ControlNet-based methods in audio synthesis quality, audio-video semantic matching, audio-video synchronization, and inference speed, pushing the boundary of ControlNet-based methods for foley synthesis.
We believe the advantage of SpecMaskFoley comes from the well-trained SpecMaskGIT backbone, the use of deep video features and the effective use of these features with our ControlNet and FT-Aligner.

Last but not least, SpecMaskFoley remains competitive even compared with strong from-scratch baselines. methods. SpecMaskFoley substantially outperforms VATT \cite{liu2024vatt}, a from-scratch MaskGIT method, in most metrics in Tab. \ref{tab:res_vgg_all}. Compared with V-AURA \cite{viertola2025vaura}, SpecMaskFoley presents superior audio synthesis quality and inference speed, while maintaining the same level of audio-video synchronization. Overall, SpecMaskFoley performed similarly to Frieren \cite{wang2024frieren}, a from-scratch flow-matching method, but our method is more advantageous in metrics such as FAD, KL, IB similarity, and DeSync. 
Note that although SpecMaskFoley uses more parameters than Frieren and MMAudio, the \textit{trainable} parameter amount is only 126M, enabling \textit{single} GPU training, hence is friendly to low-resource researchers.

According to the above observation, we believe that SpecMaskFoley has substantially narrowed the gap between from-scratch and ControlNet-based methods in the field of foley synthesis.
\vspace{-0.5mm}
\subsection{Ablation Studies and Discussions}
\label{ssec:ablation_discussion}
\vspace{-0.5mm}
\begin{table}[t]
\centering
\caption{Comparison of pretrained backbone, inference steps, and CFG settings. Bold: best score. Underline: second and third best score.}
\label{tab:res_specmaskgit_specmaskfoley}
\resizebox{\linewidth}{!}{
\begin{tabular}{l|ccc|c|c|c}
\toprule
    \multirow{2}*{Method} & \multicolumn{3}{c|}{Audio Synthesis} & AV Semantic & AV Sync. & \multirow{2}*{Infer. Time (s)$\downarrow$}\\
     & FD$\downarrow$ & FAD$\downarrow$ & KL$\downarrow$ & IB Similarity$\uparrow$ & DeSync (s)$\downarrow$ & \\
\midrule
    SpecMaskGIT-\textit{16}-step \\
    \ \ \ \ w/ text prompt & 223 & 4.97 & 2.77 & 10.2 & 1.28 & \underline{0.32}\\
    \ \ \ \ w/ audio prompt & \textbf{99.8} & \textbf{0.91} & \textbf{0.95} & 22.7 & 1.22 & \underline{0.32}\\
    SpecMaskGIT-\textit{12}-step \\
    \ \ \ \ w/ text prompt & 226 & 5.07 & 2.75 & 9.9 & 1.26 & \textbf{0.27}\\
    \ \ \ \ w/ audio prompt & \underline{108} & \underline{1.01} & \underline{0.96} & 22.7 & 1.22 & \textbf{0.27}\\
\midrule
    SpecMaskFoley-\textit{12}-step (ours) & \underline{109} & \underline{1.03} & \underline{1.76} & \textbf{26.4} & \textbf{0.65} & 0.47 \\
    \ \ \ \ CFG w/o $\ell_\mathrm{text\&video}$ & 125 & 1.15 & 1.96 & \underline{24.2} & \underline{0.79} & \underline{0.32}\\ 
    \ \ \ \ CFG w/o $\ell_\mathrm{video}$ & 125 & 1.33 & 1.82 & \underline{22.8} & \underline{0.75} & \underline{0.32}\\
\bottomrule
    \end{tabular}}
\vspace{-2mm}
\end{table}
\textbf{Effectiveness of the ControlNet branch}. SpecMaskFoley is competitive as shown in Tab.\ref{tab:res_vgg_all}, but it is still unclear that to what extent the ControlNet branch has contributed to this success. Therefore, we prompt SpecMaskGIT, the audio backbone without ControlNet, to see the behavior change before and after adding the ControlNet branch.

As shown in Tab.\ref{tab:res_specmaskgit_specmaskfoley}, SpecMaskGIT largely deviates from the desired foley synthesis when prompted with the concatenated audio tags of the test set, which might have been caused by the limited capability of the text encoder in CLAP \cite{laion2023clap}. 
Results prompted by the CLAP audio features of the audio data in the test set present the upper bound of SpecMaskGIT in terms of audio synthesis, resulting in a competitive FAD score (0.91), as well as the best KL score (0.95) and fast inference speed (0.32s for 16 steps, 0.27s for 12 steps). 
Nevertheless, the audio-video semantic matching and temporal alignment remain weak as there is no video feature injected to SpecMaskGIT.
As shown in Tab.\ref{tab:res_specmaskgit_specmaskfoley}, our ControlNet branch effectively improved the audio-video semantic matching and temporal synchronization of the backbone, without significantly degrading the audio synthesis quality.

\textbf{Impact of Multi-CFG.} It is revealed in Tab.\ref{tab:res_specmaskgit_specmaskfoley} that Multi-CFG improves all metrics compared with standard CFG, with only a slightly increased, while still affordable inference cost.

\begin{figure}[tb]
    \centering
    \begin{subfigure}[b]{0.49\columnwidth}
        \centering
        \includegraphics[width=\textwidth]{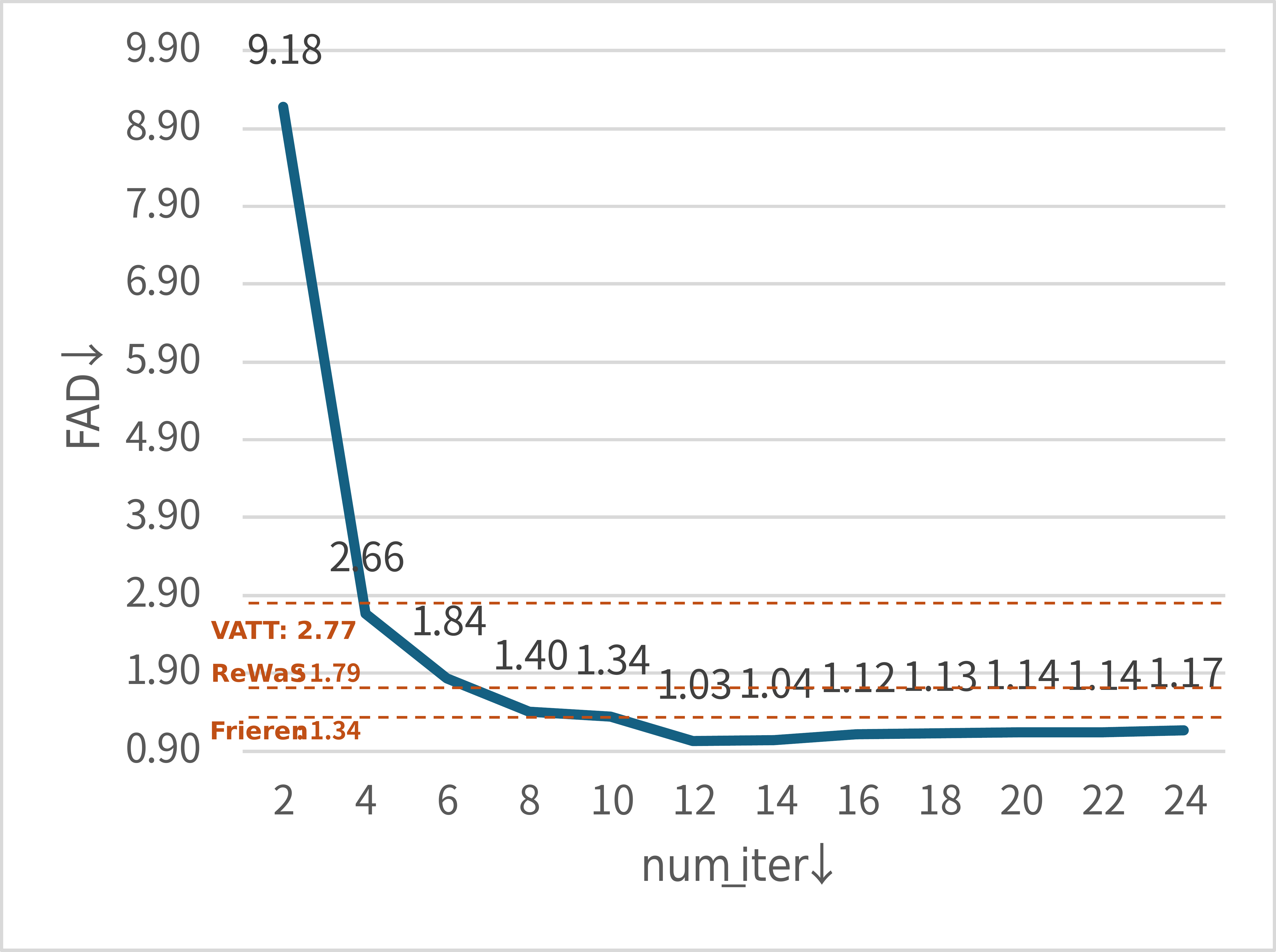}
    \end{subfigure}%
    \begin{subfigure}[b]{0.49\columnwidth}
        \centering
        \includegraphics[width=\textwidth]{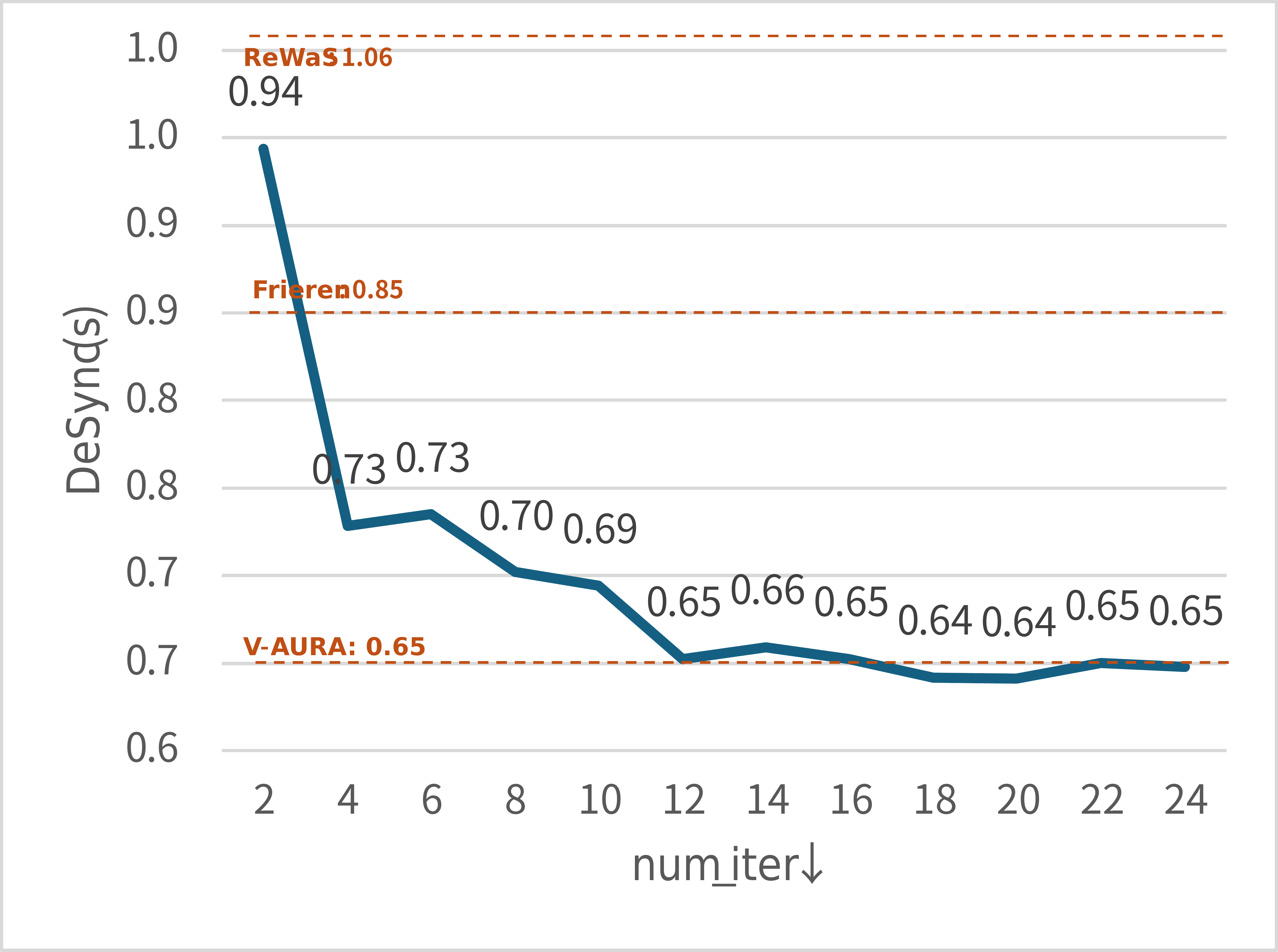}
    \end{subfigure}
    \caption{Left: FAD vs. Number of iterations. Right: DeSync vs. Number of iterations. Scores saturate after 12 iterations.}
    \label{fig:fad_desync_num_iter}
    \vspace{-2mm}
\end{figure}

\textbf{Impact of inference steps} is illustrated in Fig.\ref{fig:fad_desync_num_iter}. 
While in the original SpecMaskGIT paper \cite{comunita2024specmaskgit}, the optimal number of inference steps was 16, 
with SpecMaskFoley, we found FAD and DeSync scores saturate after 12 steps. 
It is worth noting that, using as few as \textit{4} steps, SpecMaskFoley outperforms VATT and FoleyCrafter in terms of FAD and DeSync; Using \textit{6} steps, the FAD score becomes close to that of ReWaS.
Fig.\ref{fig:fad_desync_num_iter} reveals SpecMaskFoley's few-step synthesis capability \textit{without} any distillation.

\textbf{Discussion: 2-D vs. 1-D VAE.}
As illustrated in Fig.\ref{fig:SpecMaskFoley} and discussed in Sec.\ref{ssec:implementation_details}, due to the use of a 2-D VAE in SpecMaskFoley, there are only 53 temporal frames in the latent space presenting a 10-second clip. However, the 1-D VAE used in Frieren and MMAudio presents a 10-second clip with more than 300 frames \cite{huang2023makeanaudio2}, preserving higher temporal resolution. 
According to the DeSync scores in Tab.\ref{tab:res_vgg_all}, this low temporal resolution has not been a bottleneck for SpecMaskFoley in the VGGSound benchmark, but may limit the potential of 2-D methods in the future. We leave the task of transplanting SpecMaskGIT and SpecMaskFoley to 1-D as our future work.
\section{Conclusion}
\label{sec:conclusion}
We addressed the challenging task of foley synthesis in this work.
To avoid the non-trivial task of training audio generative models from scratch, it is attractive to use pretrained TTA models. 
Many previous ControlNet-based foley synthesis methods have limited themselves to handcrafted human-readable temporal conditions.
On the other hand, from-scratch models achieved success by leveraging high-dimensional deep features extracted from pretrained video encoders.
To narrow the performance gap between ControlNet-based and from-scratch foley models, we proposed SpecMaskFoley.
Our method steers the pretrained SpecMaskGIT model toward video-synchronized foley synthesis by directly processing deep video features with ControlNet.
A frequency-aware temporal feature aligner was introduced to resolve the discrepancy between the temporal video features and the time-frequency nature of the pretrained SpecMaskGIT, simplifying the conditioning mechanism in SpecMaskFoley. 
Evaluations on a common foley synthesis benchmark demonstrate that SpecMaskFoley could even outperform strong from-scratch baselines.
Future work includes transplanting our methodology to 1-D audio MaskGIT models, and extending the scope of audio ControlNet toward spatial audio.

\clearpage
\bibliographystyle{IEEEtran}
\bibliography{refs25}
\end{document}